\documentstyle[epsf]{mn}

\def \cd {d$^{-1}$}

\def\ppm {$\pm$}
\title[HR 2740]{Discovery and analysis of Gamma Doradus type pulsations in
the F0~IV star HR 2740$\equiv$QW Pup}
\author[E. Poretti, C. Koen, P. Martinez, F. Breuer, D. de Alwis, H. Haupt]
{E. Poretti$^{1}$, C. Koen$^{2}$, P. Martinez$^{2}$, F. Breuer$^{3}$, D. de
 Alwis$^{3,4}$, H. Haupt$^{3}$ \\
$^{1}$Osservatorio Astronomico di Brera, Via Bianchi 46,
I-22055 Merate, Italy\\
$^{2}$South African Astronomical Observatory, PO Box 9, Observatory 7935,
Cape, South Africa\\
$^{3}$Summer Student at the South African Astronomical Observatory\\
$^{4}$Arthur C. Clarke Centre for Modern Technologies, Moratuwa, Sri Lanka
}

\date{ }
\begin{document}

\maketitle
\begin{abstract}
We present multi-site photometric observations of 
the F0~IV star HR 2740$\equiv$QW Pup which reveal it to be a $\gamma$ Dor
type variable pulsating with  four frequencies: 
1.0434, 0.9951, 1.1088, 0.9019 \cd. These data were obtained at the 
European Southern Observatory and South African
Astronomical Observatory over a time baseline spanning from 1997 January
14 to 1997 February 11. The 1.0434 \cd term dominates in amplitude
(10 mmag) over the other three (each less than 5 mmag); the light curve 
comprising these four frequencies 
seems to be very stable and no residual power is left in the power spectrum.
During the analysis particular attention was paid to methodological aspects,
which cannot be neglected considering the proximity of the frequencies to
1 \cd.

Physical parameters were also derived for all the well--known $\gamma$ Dor
stars, confirming that this class is very homogenous. In the framework of the
campaign, two Ap stars (OU Pup$\equiv$HR 2746 and PR Pup$\equiv$HR 2761)
were also observed. The photometric differences between these rotating
variables and HR 2740 are emphasized, corroborating the pulsational
nature of the $\gamma$ Dor stars. It is further 
demonstrated that the rotational splitting cannot be a suitable
explanation of the observed frequency content of HR~2740.

\end{abstract}
\section{Introduction}
The number of variable stars located at or near the cool border of the 
classical instability strip has rapidly increased in the recent years; they show 
complex light curves, generated by the superposition of many periods,
ranging from a few hours to a few days. Balona et al. (1994) supplied
observational evidence that $\gamma$ Doradus is a pulsating variable.
 Poretti et al. (1996; a more detailed
paper is in preparation) found a large spread in the frequencies observed
in the light curve of HD 224945, corroborating the pulsational hypothesis.
The existence of a new class of pulsating 
variable stars of around spectral type F0, that is redward of the cool
border of the classical instability strip, is now accepted and $\gamma$
Doradus has been designated the prototype of this class. These stars have
periods on order of 1 d and $V$ amplitudes $\sim$0.02 mag.

Since $\gamma$ Doradus stars are located in a 
region of the HR diagram where variable stars are not expected, they were
often used as comparison stars; a revisitation
of old photometric measurements was undertaken in order to detect new
members. In such a  way,  Breger et al. (1997) suggested that the low frequency
content
of the light variability of the $\delta$ Sct star 4 CVn must be ascribed to
its comparison star, the F0 star HD 108100.

The light variability of the F0~IV star HR2740$\equiv$HD 55892$\equiv$QW Pup
 was discovered
by Hensberge et al. (1981), who used it as a comparison to measure some
Ap stars. On the basis of 20 $uvby$ data collected on 16 nights spanning 
41 days, they derived a period of 0.9363$\pm$0.005 d and considered HR 2740 as
a ``mild Ap star". However, Hensberge et al. also emphasized the lack of
the continuum depression near 520 nm (typical for Ap stars) and the necessity
to perform additional photometry to confirm period and classification.

The link between HR 2740 and the $\gamma$ Doradus stars was definitely 
established when when we took some high resolution spectra of HR 2740 
(considered by Slettebak et al. 1975 as as standard for rotational velocity)
and clearly detected some line profile variations. 
Now that many F0 stars are known to be variable and are grouped into a
new class, it is quite obvious to realize that Hensberge et al. (1981) 
performed a pre--discovery observation of a $\gamma$ Dor variable star.

\section{Photometric observations}
As demonstrated by the previous experiences on other $\gamma$ Dor stars,
single--site observations are not suitable to study their light variability.
The aliases at $\pm$1\cd are strong, multiperiodicity is commonly observed
and the signal is concentrated near 1 \cd: all these
factors produce complex spectral patterns, where the detection of the true
frequencies is a difficult exercise (see for example Mantegazza et al. 1994).
For these reasons, new observations of HR 2740 were obtained from two sites:
at La Silla (European Southern Observatory, Chile) and Sutherland (South African
Astronomical Observatory, South Africa). Unfortunately a third observing
site located in New Zealand could not join the project.

Moreover, it was decided to add the Ap stars HR 2761$\equiv$PR Pup and 
HR 2746$\equiv$ OU Pup to the observing
programme: these variables are known to have periods similar in length to
that expected for HR 2740, a small amplitude and well defined light
curve. Located near HR 2740, PR and OU Pup provide an excellent litmus--paper
to compare and check the results obtained on HR 2740. HR 2762 and HR 2789
were selected as comparison stars; since all the stars are
bright, the Str\"omgren filters were preferred to the Johnson ones.

The 50-cm telescope  equipped with a EMI 9789QA tube was used at ESO (observer
E.~Poretti). The measurements were performed on eight consecutive nights,
from 1997 January 29--30 to 1997 February 5--6. The frequency of measures was 
different for the different stars: since the variability of the two Ap stars was known to be slow
and monoperiodic, HR 2740 was measured (in $uvby$) 6 times more frequently.

A 50-cm telescope equipped with a Hamamatsu tube was also used at SAAO; however,
to avoid saturation effects the telescope was stopped out at an aperture of
28 cm.
A first string of data covers 6 nights of 7 from 1997 January 14--15
to 20--21 (observers P.~Martinez, D.~de Alwis, F.~Breuer, H. Haupt); then a second string 
of data covers 10 nights
from 1997 January 28--29 to 1997 February 10--11 (observers C.~Koen and P.~Martinez).
Measurements were performed in $b$ and $y$ light and the two Ap stars were
measured more frequently than at ESO.
\section{Data reduction}
\subsection{Differential photometry}
The data reduction was performed following the method described by Poretti
\& Zerbi (1993), which allows us to determine the behaviour of the
extinction coefficient $k_\lambda(t)$ during the night.
Table 1 summarizes the results of the behaviour of  $k_\lambda$. 

Since the measurements were generally performed in excellent photometric
conditions, the very small fluctuations observed in the $k_\lambda(t)$ values
(upper part of Tab. 1) and the close proximity of all the stars
did not introduce any appreciable change (a very few measurements changed by 
1 or 2 mmag) with respect to the magnitude
differences calculated by the usual Bouguer's line (i.e. $k_\lambda$
considered as a constant value). In turn, this comparison
confirms the stability of a photon counting system properly controlled 
(i.e. no instrumental drift from one night to the next) and
the colour--dependant variation of the extinction coefficient (lower part
of Tab.1) strongly supports the atmospheric origin of the variation.
Changes in transparency were occasionally observed when the sky became clear
after a period of cloudy weather (see lower part of Tab. 1 and Fig. 1)
\begin{figure}
\epsfxsize=8.5truecm
\epsfysize=11.5truecm
\centerline{\epsffile{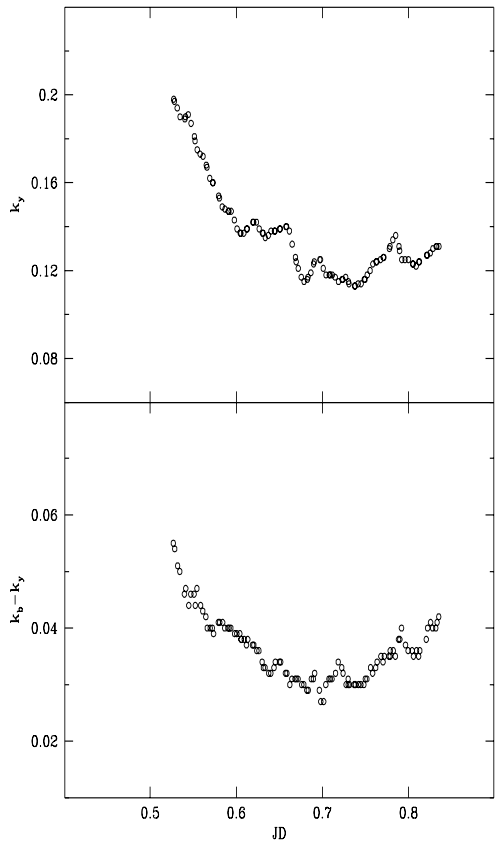}}
\caption[]{The variability of the exctinction coefficient on night
JD 2450480 as observed at ESO; the behaviour depends on the
 wavelength, as showed by the differences between the $k_b$ and $k_y$ values}
\end{figure}
\begin{table}
\centering
\caption[]{Extremum values for the extinction coefficient. Units are 
mag~airmass$^{-1}$}
\begin{tabular}{cccrcrc}
\hline
\hline
Filter&& &\multicolumn{1}{c}{ESO}  & &&\multicolumn{1}{c}{SAAO}\\
&& &Min--Max& && Min--Max\\
\cline{4-4} \cline{7-7} 
\multicolumn{7}{l}{All the nights}\\
$u$&& & 0.49--0.70 & && -- \\
$v$&& & 0.27--0.44 & && -- \\
$b$&& & 0.11--0.26 & && 0.14--0.21 \\
$y$&& & 0.09--0.20 & && 0.09--0.15 \\
\noalign{\smallskip}
\multicolumn{7}{l}{Maximum range in one night}\\
\noalign{\smallskip}
$u$&& & 0.52--0.70 & && -- \\
$v$&& & 0.30--0.44 & && -- \\
$b$&& & 0.14--0.25 & && 0.16--0.20 \\
$y$&& & 0.11--0.20 & && 0.10--0.15 \\
   && & \multicolumn{1}{c}{JD 2450480}& && \multicolumn{1}{c}{JD 2450490}\\
\hline
\hline
\end{tabular}
\end{table}

For sake of homogeinity, the same algorithm was used both
for ESO and SAAO measurements to calculate the differential magnitudes of all
the stars respect with HR 2762. No  variability was found in the light curves
of HR 2789, i.e. both the comparison stars can be considered constant in
brightness. As regards the 370 ESO measurements, the rms residuals are 3.9, 3.2,
3.3 and 3.5 mmag in the $uvby$ colours, respectively;  as regards the 265 SAAO 
measurements they are 3.1 and 2.9 mmag in the $by$ colours, respectively. After
alignment
of the mean magnitudes, the resulting residual rms's are 3.2 and 3.3 mmag
in the $by$ colours, respectively.

\subsection{Mean magnitudes alignments}
The determination of the systematic difference in the zeropoints constitutes
 a serious problem when dealing
with multiperiodicities near 1 d and different instrumental systems. 
Let us discuss the matter starting from the differential measurements of the
 two Ap stars,
which are monoperiodic variables (see also the next section). In this case,
once the period is determined by analyzing each dataset, it is possible to
to obtain the two zeropoints simply by performing a least--squares fit. 

In the case
of OU Pup, the SAAO and ESO datasets each have very good phase 
coverage and they can be used separately to yield two sets of parameters; 
it was then
possible to verify that amplitudes and phases agreed very well and the
difference between the two zeropoints immediately yielded the systematic
shift in magnitude. In the case of PR Pup, only the SAAO dataset covers
in phase the whole light curve; therefore a different procedure was followed.
From the fit of SAAO data we calculated the frequency and amplitude of the
sine--wave and then
we fitted the ESO data by keeping these values locked. In this way we
avoided
obtaining unreliable values for the amplitude, since the extrema of the light
curve not well covered in the ESO data. From this locked fit
the ESO zeropoint was also derived and the two datasets could be  merged.

The case of HR 2740 is quite different. As described above, multisite
observations are necessary to decipher the light curve and hence no
preliminary
solution can be reliably used. The same problem was recently discussed by
Breger et al. (1997; see their Sect. 2); those authors decided to subtract mean
values from each dataset and then merge the data. This procedure is
particularly dangerous in the case of $\gamma$ Dor stars since most of the
signal is concentrated near 1 \cd and then and a  considerable reduction 
in amplitude can
occur. This can be easily verified in the case of OU Pup, where the light
curves over the period 0.918 d were obtained separately for each dataset: the 
shifts so calculated are 1.8 and 2.1 mmag in $b$ and $y$ light; on the other
hand, the mean magnitudes are shifted by 4.1 and 3.9 mmag, respectively.

 Therefore we decided to calculate the systematic shifts between the two
datasets
by using the overlapping segments of the light curve; the end of
the observations at Sutherland coincides with the start at La Silla. As
noted by Breger et al. (1997), measures done at large airmasses are not very
accurate. However this method can provide satisfactory results; for example,
starting from an error of about 3 mmag on a single observation (as
in our case) and admitting that this error may be doubled at large airmass,
if we have 4 measurements for each site performed in the overlap time, then 
in one night the systematic shift can be calculated with a precision of about
4.2 mmag and  5 nights are sufficient to reduce the error on the shift to
less than 2.0 mmag. Of course, it is sufficient to have 9 nights or 8
measurements per
site to ensure an error of about 1 mmag. So, the potential of the overlap
method should be carefully considered in the planning and in the data 
reduction of multisite
campaigns. In the case of HR 2740, we applied this method and determined a
systematic shift of
13 mmag in $b$~light and 6 mmag in $y$~light; these values are slightly
different from those obtained by aligning the arithmetic average of the two
datasets (17 and 10 mmag, respectively). From a close inspection
of the light curves it seems that the alignments produced by the 
values supplied by the average method are not satisfactory;
we will discuss this point again at the end of next section.

\section{The frequency analysis}
To perform the frequency analysis we used the Vani\^cek's (1971)
least-squares technique, which searches for  multiple periods without relying
on prewhitening, thus avoiding dangerous deformations in the power
distribution. The `reduction factor'
 RF=1--$\sigma^2_{\rm fin}/\sigma^2_{\rm in}$  is
calculated for each trial frequency. $\sigma^2_{\rm in}$ is
the variance before considering it and 
$\sigma^2_{\rm fin}$ is the variance after considering it; if the RF is close
to 1 it means that the frequency fits the data very well, greatly reducing
the variance, while if RF is near 0 this means that the frequency doesn't
fit the data appreciably.

The parameters of the least--squares fit were calculated by fitting the data
to the series
\begin{equation}
\Delta m(t)  =  \Delta m_0 + \sum_{j=1}^M A_j \cos [2\pi f_j
 (t-T_0) + \phi_{j}] 
\end{equation}
\subsection{OU Pup}
The frequency analysis of the ESO and SAAO measurements were performed
separately. Both yielded 1.09 \cd as the
highest peak in the power spectrum; introducing it as the ``known
constituent" (k.c.)
 the 2$f$ harmonic 
was detected. These two terms strongly reduced the variance
and no further frequency components could be detected. The measurements were
merged by means of the procedure described above. We confirm the
period value given by Heck et al. (1987), i.e. 0.9189 d; the
mean light curves are shown in Fig. 2 and are characterized by a constant
brightness for about 0.35\% of the period.
The simplest explanation of such
a variability is to think of a bright feature on the stellar surface carried 
through the visible disk by
rotation; this feature is not always seen 
from the Earth and its disappearance produces the standstill at 
minimum brightness.  As can be noted (see Fig. 2 and Tab. 2),
 the $u$ light curve 
has an amplitude much larger than in the other colours; in particular,
in $y$~light the curve is very shallow.

\begin{figure}
\epsfxsize=8.5truecm
\epsfysize=11.5truecm
\centerline{\epsffile{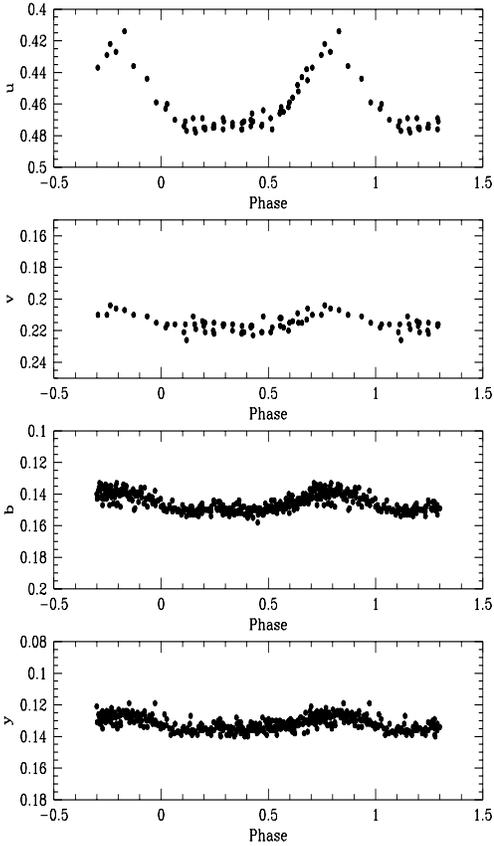}}
\caption[]{$uvby$ light curves of OU Pup. $b$ and $y$ measurements were
performed both at ESO and SAAO, while $u$ and $v$ data were obtained only at ESO}
\end{figure}
\subsection{PR Pup}
The frequency analysis supplied only one peak at 0.485 \cd and no harmonic.
This value for the frequency is particularly unfavourable and hampered any
direct calculation of  reliable $A_0$ values. However, the procedure described
in the previous  section allowed us to obtain the mean light curves
shown in Fig. 3.
In the case of PR Pup the light curve is perfectly sinusoidal, implying 
that the feature responsible for the variability is always visible
from the Earth; also in this case the largest amplitude is observed in
$u$ light.

There is a bit of confusion about the period of this star in the literature:
Heck et al. (1987) report 1.9346 d (i.e. 0.517 \cd), Catalano \& Leone (1993)
report 2.06370 d (i.e. 0.485 \cd). It is evident that one period is the alias
of the other; our double--site observations allowed us to exclude the former
and to confirm the latter.
We determined a time of minimum brightness at HJD 2450464.59, 
with an O--C of --0.47 d respect with the ephemeris reported by Catalano \& 
Leone (1993).
If we admit no period variation and no phase shift, a refined period of
2.063243 d can be derived by combining the two times of minimum light.
The parameters of the least--squares fit are reported in Tab. 2.
\begin{table*}
\centering
\caption[]{Least--squares fit of the data on PR Pup and OU Pup}
\begin{tabular}{l c rr c rr c rr c rr }
\hline
\noalign{\smallskip}
      & & \multicolumn{2}{c}{$u$}& &\multicolumn{2}{c}{$v$}
      & &\multicolumn{2}{c}{$b$} & &\multicolumn{2}{c}{$y$}\\
\cline{3-4} \cline{6-7} \cline{9-10} \cline{12-13}
Star& & Ampl. & Phase & & Ampl. & Phase & & Ampl. & Phase & & Ampl. & Phase \\
    & & [mmag]& [rad] & & [mmag]& [rad] & & [mmag]& [rad] & & [mmag]& [rad] \\
\noalign{\smallskip}
\hline
\hline
\noalign{\smallskip}
OU~Pup    &&   23.5&    4.43  &&  5.2 &   4.41 &&   5.4 &    4.36 &&   4.0 &   4.31\\
$f$=1.0889 \cd&&\ppm5.2&\ppm0.54&&\ppm0.7&\ppm0.13&&\ppm0.4 &\ppm0.12&&\ppm0.4&\ppm0.14\\
\noalign{\smallskip}
2$f$      &&    9.4&    5.58  &&  2.4 &   5.93 &&   2.1 &    5.69 &&   1.7 &   5.07\\
          &&\ppm5.5&\ppm0.59&&\ppm0.7&\ppm0.26&&\ppm0.5 &\ppm0.26&&\ppm0.4&\ppm0.33\\
\noalign{\smallskip}
 &  \multicolumn{11}{c}{$T_0$~=~Hel. J.D. 2450463.000} \\
$\Delta m_o$ & &\multicolumn{2}{c}{0.4575\ppm0.0006}& &\multicolumn{2}{c}{0.2148\ppm0.0005}
 & &\multicolumn{2}{c}{0.1465\ppm0.0002}& &\multicolumn{2}{c}{0.1325\ppm0.0002}\\
Residual rms & &\multicolumn{2}{c}{3.7 mmag}
 & &\multicolumn{2}{c}{3.2 mmag}
 & &\multicolumn{2}{c}{2.9 mmag}
 & &\multicolumn{2}{c}{2.9 mmag}
 \\
\hline
PR~Pup    &&   56.6&    1.60  &&  14.8 &   1.37 &&   16.2 &    1.38 &&  17.6 &   1.43\\
$f$=0.4846 \cd&&\ppm4.4&\ppm0.14&&\ppm1.6&\ppm0.32&&\ppm0.3 &\ppm0.05&&\ppm0.3&\ppm0.05\\
\noalign{\smallskip}
 &  \multicolumn{11}{c}{$T_0$~=~Hel. J.D. 2450463.000} \\
$\Delta m_0$ & &\multicolumn{2}{c}{0.7292\ppm0.0007}& &\multicolumn{2}{c}{0.9558\ppm0.0004}
 & &\multicolumn{2}{c}{0.9663\ppm0.0002}& &\multicolumn{2}{c}{0.9740\ppm0.0002}\\
Residual rms & &\multicolumn{2}{c}{4.8 mmag}
 & &\multicolumn{2}{c}{3.0 mmag}
 & &\multicolumn{2}{c}{3.2 mmag}
 & &\multicolumn{2}{c}{3.1 mmag}
 \\
\hline
\hline
\end{tabular}
\end{table*}

\begin{figure}
\epsfxsize=8.5truecm
\epsfysize=11.5truecm
\centerline{\epsffile{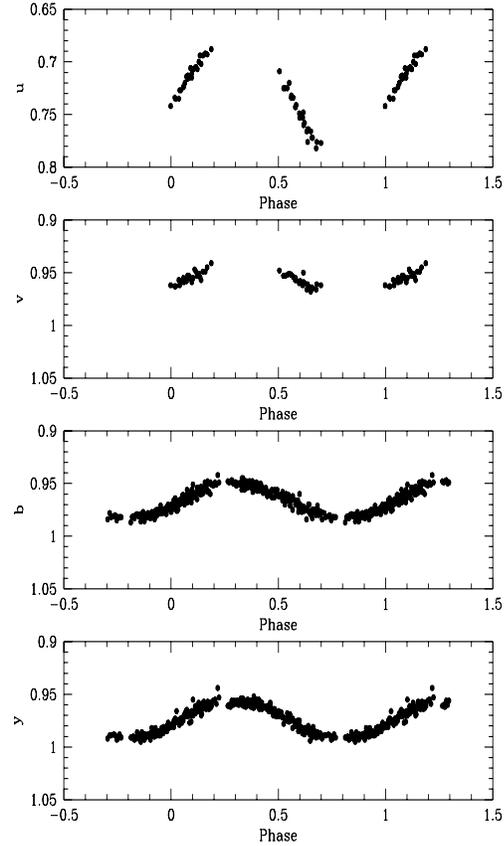}}
\caption[]{$uvby$ light curves of PR Pup. The unsatisfactory phase coverage
of the single--site measurements is shown by the $u$ and $v$ light curves
(ESO only); while $b$ and $y$ measurements were performed both at ESO and
SAAO}
\end{figure}

\subsection{HR 2740}
\begin{figure*}
\epsfxsize=15.5truecm
\epsfysize=11.5truecm
\centerline{\epsffile{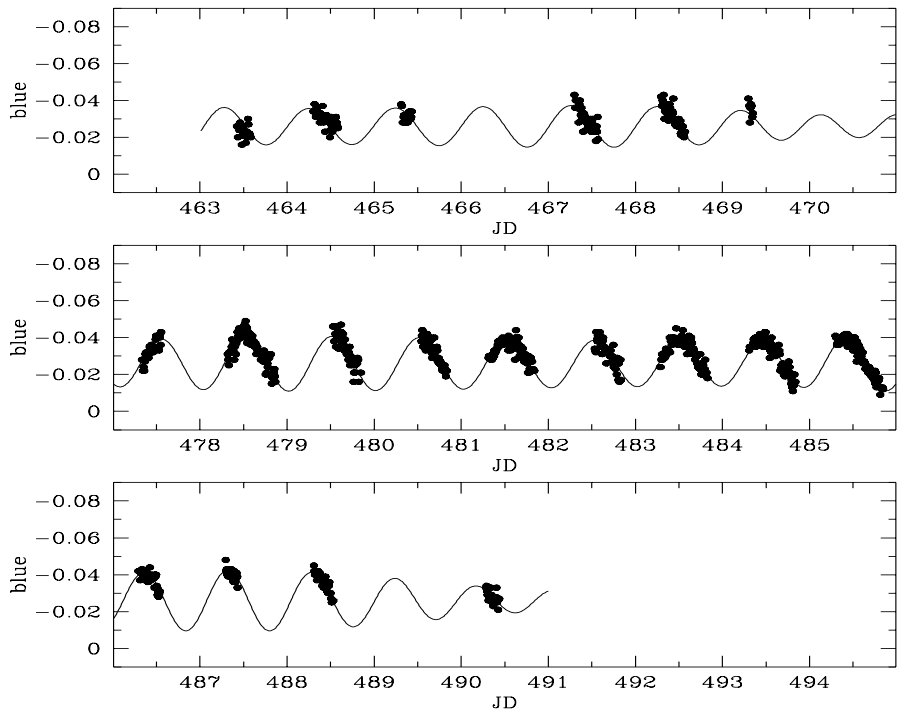}}
\caption[]{$b$ light curves of HR 2740.}
\end{figure*}
\begin{figure*}
\epsfxsize=15.0truecm
\epsfysize=13.5truecm
\centerline{\epsffile{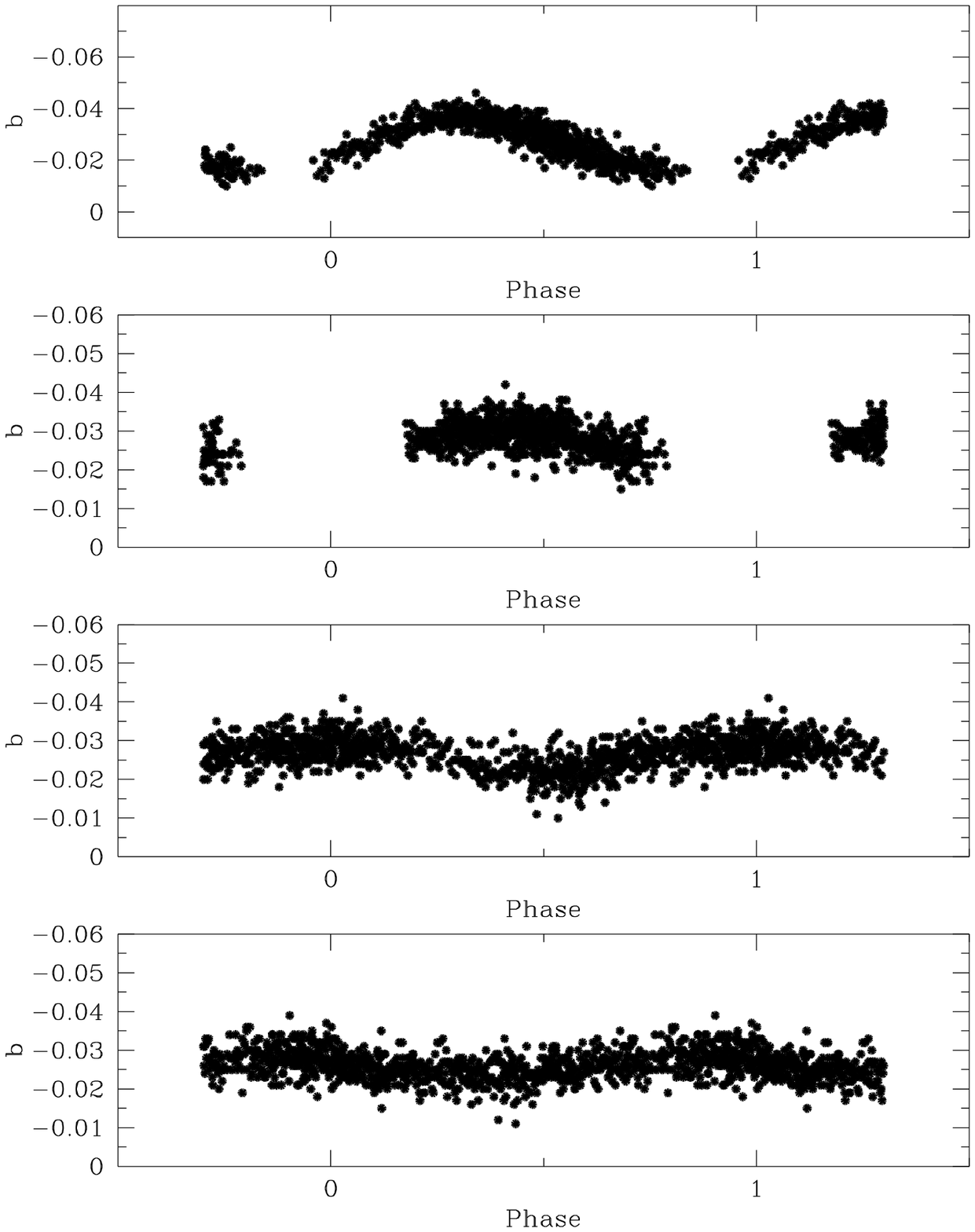}}
\caption[]{Power spectra of the $b$ light curves of HR 2740. From top to bottom:
with no known constituent (k.c.), with $f_1$=1.04 \cd as k.c.; with $f_1$ and
$f_2$=1.10 \cd as k.c.'s; with $f_1$, f$_2$, $f_3$=0.90 \cd as k.c.'s;
with $f_1$, f$_2$, $f_3$,  $f_4$=0.995 \cd as k.c.'s}
\end{figure*}

Figure 4 shows the light curve in $b$; in the intervals 
JD 2450463--2450470 and JD 2450486--2450491 we have only measurements carried
 out at SAAO (upper and lower panel), while
in the interval JD 2450477--2450486 we have contiguous 
measurements carried out at SAAO and ESO (middle panel).

Both $b$ and $y$ data were analyzed in frequency and we discuss 
here the former,
 since in that colour the amplitudes are larger.
A simple look at the middle panel of Fig. 4 clearly suggests a period
close to 1 day and a regular light curve. The ESO measurements 
always show a regular decrease in magnitude, perfectly linked to the SAAO ones
which end at maximum brightness. However, the SAAO light curve
at the beginning of the campaign is less regular and does not
seem to repeat itself regularly from one night to the next. 
We firstly considered the dataset comprising the measurements obtained
from JD 2450477 to JD 2450486 only, since the spectral window of this dataset
is relatively free from the 1 \cd alias and this allows more reliable
frequency identifications. Indeed 
we obtained a well defined peak at $f_1$=1.02 \cd, explaining
the variation evidenced in the middle panel of Fig. 4. To proceed further, we
included it as a known constituent (k.c.) in the successive least--squares 
search; this means that considering Eq. (1) applied to $b$ data,
the unknowns in the second search were $\Delta b_0, A_1, \phi_1, f_2, A_2, 
\phi_2$, while the $f_1$=1.02 \cd value was kept fixed. As a second peak
we found $f_2$=0.90 \cd; the fit with $f_1$ and $f_2$ leaves a 
residual rms of 3.1 mmag and considering that this value is very close
to the  standard deviation of the measurements of HR 2762 we could
conclude that  these two frequencies yield a good explanation of the observed
light curve; only small residual peaks are visible in the spectrum obtained
by introducing $f_1$ and $f_2$ as k.c. It must be emphasized that the power
spectra are absolutely flat after 4 \cd; the signal is entirely contained
in the low frequency range.

On the basis of the above analysis, the light variation of HR 2740 seems to
be well described by one dominant frequency and one or two low-amplitude
terms. However, the analysis of the measurements
performed at SAAO in the interval between JD 2450463 and 2450470 distrupted
this simple scenario. When considering the solution just obtained, we were
not able to fit this part of the light curve in a satisfactory way.
Since the rising branch was observed in the
JD 2450478--2450487 interval and the gap was only 8 days, the phases on
the $f_1=1.02$ \cd cycle should be the only slightly (less than 0.2 period)
shifted
towards the minimum light branch in the JD 2450463--2450470 interval, but this
 was not observed. Also the measurements carried out in the JD 
2450486--2450491 interval could not be fitted in a satisfactory way; in
particular, the descending branch observed in the last night at SAAO
is too steep and, overall, too early in phase to be linked to the previous ones
on the basis of the $f_1$=1.02 \cd term.

The light variation of HR 2740 is too complex to be accounted for by a
single $f_1=1.02$~\cd oscillation. Hence,
we analyzed the whole set of data; its spectral window shows
a higher 1 \cd alias, but the frequency resolution is greatly improved by
the more extended time baseline (from 9 to 27 days).
The upper panel of Fig. 5 reveals the true frequency content. The previously
 detected peak at 1.02 \cd is a
blend of two close peaks. In the upper panel the highest peak is at 1.04 \cd,
and the close peak at 0.99 is the highest in the fourth panel. These two terms
generated a beat phenomenon.
In the JD 2450478--2450487 interval they are in phase coherence, while 
at the beginning of the observations the shape of the light curve was
deformed by the phase oppositions of the two components. In addition, two
other terms can be detected, at 1.10 \cd (second panel) and around 0.90 \cd
(third panel; it is the same term detected in the previous analysis). Since one
of them could be the alias  of the other (0.90=1.00-0.10, 1.10=1.10+0.10), we 
checked whether only one of them can explain the light variation
by omitting one term in the least--squares solution or by changing the order
of introduction of k.c. in the frequency analysis; we
always  found both in the power spectra. The interaction of one term with the
alias of the other enhanced the power of the peak at 1.10 \cd, which is
detected as the second term even if its amplitude is smaller than that of
the 0.99 \cd term. We also verified our solution by 
introducing $f$=1.0434 \cd and $f$=0.9951 \cd as k.c., but again the 
least--squares method found $f$=1.108 \cd as third component and $f$=0.908
\cd as the fourth.

 When introducing these four
frequencies as k.c., we obtained  a rather flat spectrum also in the low
frequency region (lower panel).
If other small amplitude terms are present, they have an amplitude lower than
1 mmag and cannot be detected in a reliable way.
We also investigated the possibility that the signal has a 
non--sinusoidal shape by fitting the sum of a fundamental frequency
and several harmonics. Such a procedure give a best-fit frequency of 1.02
\cd.
However, the residuals are not white and the removal of several further
 sinusoids is 
necessary before the spectrum is as flat as in the bottom panel of Fig. 5,
clearly supporting a multiperiodic content.

The presence of a term very close to 1 \cd can be considered as suspect,
and is possibly due to a misalignement between the two
datasets. Firstly, it must be noted that a term close to 1.04 \cd is 
necessary to explain the beating in the light curve. In any case,  the most direct
way to check its physical
nature is to process the dataset comprising the SAAO measurements only.
This dataset is affected by the strength of the alias at $\pm$1~\cd, but
spanning the whole length of the campaign it is adequate to resolve the two
peaks, a possibility that the shorter ESO dataset does not offer. Its analysis
reveals that, even
if the alias at 2.04 \cd is as high as the term at 1.04 \cd and the 1.10
\cd term diverts power to the 0.90 \cd one, the
term at 0.99 \cd is visible in the power spectrum. It is important to note
that the analysis of a subset yields us the same frequencies as the whole
set; this supports the working idea that the four terms are really present in
the light curve and are not generated by one or two terms showing amplitude
and/or phase variations, since in such a case the frequency content should
change when reducing the sample.

Moreover, we performed additional tests changing the value of the correction in
magnitude applied to the SAAO measurements; in all the cases the signal at
0.99 \cd is always present. We  also noted that the rms residual has
the minimum value for the adopted shift (13 mmag), which was calculated from overlapping
segments of the light curves and hence is independent of the solution. Moreover,
once the accepted solution was obtained, we performed several fits on the
two datasets keeping locked some of the parameters and evaluating the 
differences between the zeropoints. We always obtained a value around 13 mmag, 
never reaching the 17 mmag value yielded by the average method. Even if
the adoption of such an extreme value did not prejudice the determination
of the solution, it should be regarded as an artifact of the odd distribution
in phase.  When considering the discrepancy between the two shifts
calculated in the OU Pup case (see Sect. 2.2), we obtain a serious warning
against the use of the mean level subtraction as a method to correct
systematic shifts.

Hence, we can be very confident of the physical nature of the 0.99 \cd
term. It must be noted that its amplitude is very similar in $b$ and $y$ light,
while in the other cases the $y$ amplitude is about  20\% smaller 
than the $b$ amplitude. Even if error bars on the amplitudes 
can account for this discrepancy,
it is possible that this term has a geometrical original, e.g. 
an ellipsoidicity effect due to a close companion.

Table 3 lists the coefficients of the least--squares fit. Note that
in the $b-y$ data only the $f_1$=1.04 \cd term was
considered. Of course, the other terms are
present in the $b-y$ light curve since their amplitudes in $b$ and $y$ 
are different, but since the amplitude differences are less than 1 mmag 
their consideration changes the rms of the residuyals marginally and their least--squares
parameters are not useful, being affected by large uncertainties.
Unfortunately, since $u$ and $v$ measurements were performed at ESO only, 
the related datasets did not have the necessary resolution to yield a reliable
solution which separates the contributions of the two close terms at 1.04 and
0.99 \cd.

We also analyzed the time series by means of the {\sc clean} technique (Roberts
et al. 1987), but, as  already noted in the case of HD 224845 (Mantegazza et
al. 1994), this method is not suitable for frequencies near 1 \cd since 
the subtraction of the average before computing the periodogram  removes not
only the power at zero frequency, but also most of it at its strong aliases
(1.0 \cd, 2.0 \cd, ...).

\begin{table*}
\centering
\caption[]{Least--squares fit of the data on HR 2740}
\begin{tabular}{r c rr c rr c rr }
\hline
\noalign{\smallskip}
      & & \multicolumn{2}{c}{$blue$}& &\multicolumn{2}{c}{$yellow$}
      & &\multicolumn{2}{c}{$b-y$}\\
\cline{3-4} \cline{6-7} \cline{9-10}
Freq. & & Ampl. & Phase & & Ampl. & Phase & & Ampl. & Phase \\
$[d^{-1}$] & & [mmag]& [rad] & & [mmag]& [rad] & & [mmag]& [rad] \\
\noalign{\smallskip}
\hline
\hline
\noalign{\smallskip}
1.0434 & &   10.5 &    1.18 & &    8.3 &    1.23 & &    1.6 &    1.22 \\
\ppm0.0004& &\ppm0.4 &\ppm0.05 & &\ppm0.3 &\ppm0.07 & &\ppm0.2 &\ppm0.23 \\
0.9951 & &    4.8 &    0.59 & &    4.6 &    0.55 & \\
\ppm0.0010 & &\ppm0.5 &\ppm0.11 & &\ppm0.5 &\ppm0.12\\
1.1088 & &    3.5 &    3.16 & &    2.8 &    3.28 \\
\ppm0.0017 & &\ppm0.2 &\ppm0.20 & &\ppm0.2 &\ppm0.24\\
0.9019 & &    2.4 &    4.14 & &    1.9 &    4.17\\
\ppm0.0020 & &\ppm0.3 &\ppm0.15 & &\ppm0.3 &\ppm0.20 \\
\noalign{\smallskip}
 &  \multicolumn{9}{c}{$T_0$~=~Hel. J.D. 2450463.0000} \\
 & &\multicolumn{2}{c}{$b_0$~=~--0.0258\ppm0.0003}& &\multicolumn{2}{c}{$y_0$~=~--0.2739\ppm0.0002}
 & &\multicolumn{2}{c}{$(b-y)_0$~=~--0.2475\ppm0.0002}\\
 & &\multicolumn{2}{c}{Residual rms 3.2 mmag}
 & &\multicolumn{2}{c}{Residual rms 3.2 mmag}
 & &\multicolumn{2}{c}{Residual rms 3.7 mmag}
 \\
 & &\multicolumn{2}{c}{878 measurements}
 & &\multicolumn{2}{c}{878 measurements}
 & &\multicolumn{2}{c}{874 measurements}
 \\
\hline
\hline
\end{tabular}
\end{table*}       
\section{Discussion}
\begin{figure}
\epsfxsize=8.5truecm
\epsfysize=11.5truecm
\centerline{\epsffile{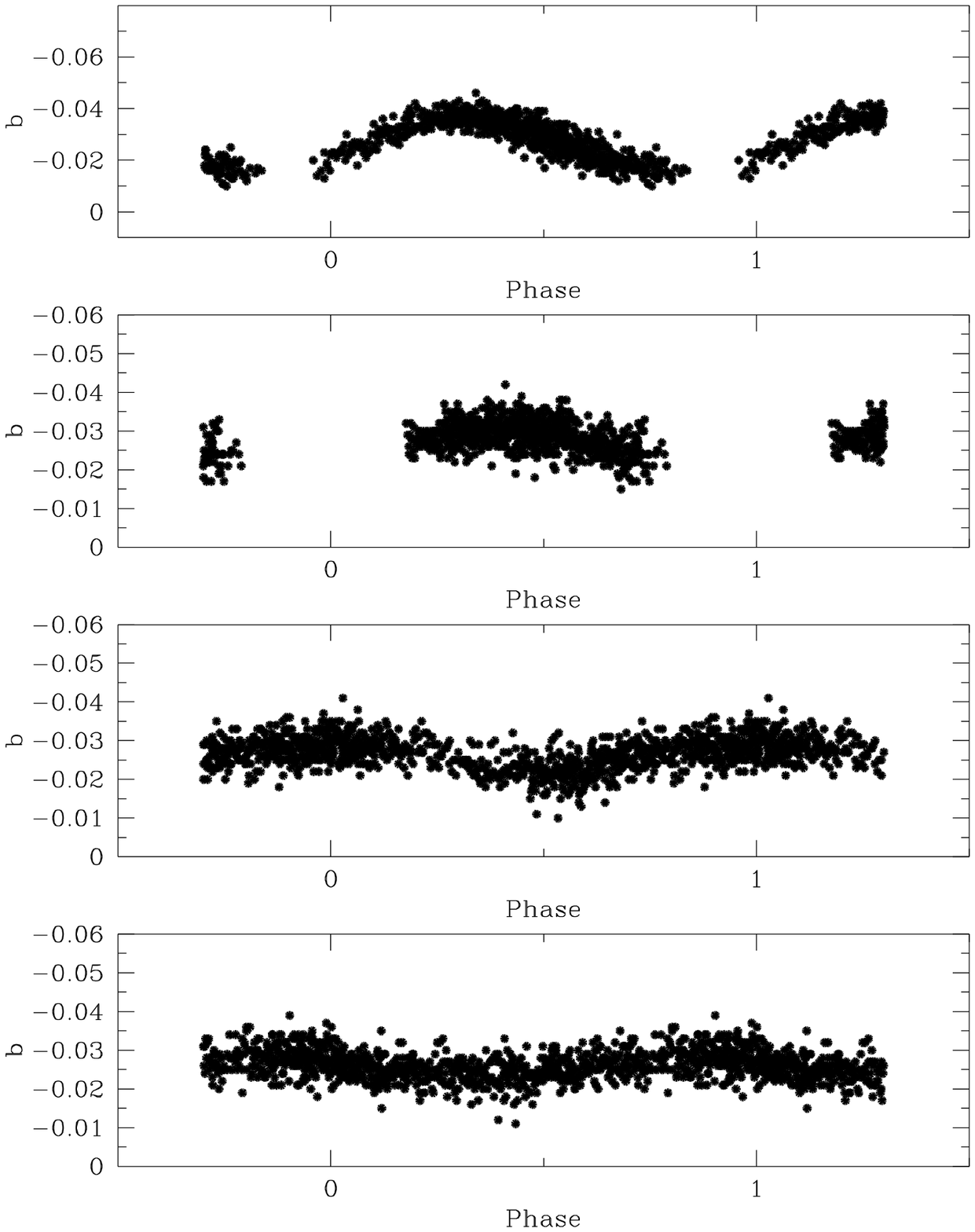}}
\caption[]{$b$ light curves of the four terms evidenced in the light curve
of HR 2740}
\end{figure}
We established that HR 2740 is a new $\gamma$ Dor variable, showing a 
dominant frequency of 1.043 \cd, beating with a term at 0.995 \cd; two
other terms at 0.902 and 1.109 \cd complete the fit of the light
curve and no other periodicities could be found in our double--site
measurements. Figure 6 shows the $b$ light curves of the four components we
identified. To obtain each of them we subtracted  the theoretical contribution (by using
the least--squares fit listed in Tab. 3) of the other three 
from the original measurements.
To search for possible long-term variations in the frequency content of the
oscillations we re-examined the photometric data of Hensberge et al. (1981),
where HR~2740 was used as a comparison star. 
 Unfortunately, those measurements  are too scanty (21
points on 16 nights spanning 41 days) to provide an additional
dataset and to yield details of long--term behaviour.

\subsection{Stellar parameters}
 In recent years, a number of multi-site observing studies have been
conducted on this new class of
variable stars and it is interesting to compare the new results obtained for
HR 2740 with the phenomenology evidenced in previous cases.
Reliable $uvby$ standard photometry is available for HR 2740:
$b-y$=0.219, $m_1$=0.154, $c_1$=0.640, $\beta$=2.705. By means of the 
Moon \& Dworetsky (1985) calibration we obtained typical values for a star located
near the cool border of the classical instability strip, i.e. $R$=1.92$R_\odot$, 
$T_{\rm eff}$=6850 K, $M_V$=2.44; moreover, the $\delta m_o$ value of 0.010
suggested normal metal abundances. Curiously, HR 2740 has the same
(within the error bars) physical properties of HD 108100; however, this
seems true of other $\gamma$ Dor stars too, as shown by the paramaters listed in Tab.4.
Since they were obtained by applying the same procedure, the narrow spread
of each parameter for different stars has a physical meaning,
suggesting us that all the $\gamma$ Dor stars are very similar objects.
\begin{table*}
\centering
\caption{$\gamma$ Dor stars ordered for increasing $b-y$: physical parameters
obtained from $uvby\beta$ photometry} 
\begin{tabular}{l cccc r rcccc }
\hline
Star & $b-y$& $m_1$ & $c_1$ & $\beta$ && $\delta m_0$ &
 $M_V$ & $R/R_\odot$ & $T_{\rm eff}$ & $\log\;g$ \\
 & & &  &  && &
       &             &  [~K~]     &       \\
\noalign{\smallskip}
\hline
\noalign{\smallskip}
HR 8799      &0.181&0.142&0.678&2.745&&  0.04&2.85&1.48&7250&4.3\\
HD 224945    &0.192&0.147&0.719&2.743&&  0.04&2.45&1.82&7200&4.1\\
HD 224638    &0.198&0.154&0.690&2.726&&  0.02&2.41&1.87&7050&4.0\\
$\gamma$ Dor &0.206&0.165&0.670&2.742&&  0.02&2.88&1.52&7200&4.2\\
9 Aur        &0.212&0.155&0.643&2.723&&  0.02&2.78&1.64&7050&4.2\\
HR 2740      &0.219&0.154&0.640&2.705&&  0.01&2.44&1.92&6850&3.9\\
HD 164615    &0.230&0.178&0.624&2.715&&--0.01&2.80&1.66&7000&4.1\\
HD 108100    &0.234&0.161&0.639&2.705&&  0.00&2.45&1.97&6850&3.9\\
\noalign{\smallskip}
\hline
\end{tabular}
\end{table*}
The quoted physical parameters can be used to calculate the pulsational 
constants $Q_i$ for the four modes observed in the light curve of HR 2740,
obtaining 0.35$<Q_i<$0.43 d, i.e. values typical for gravity modes.
The radial fundamental mode is expected to have a value of about 12 \cd;
no term having an amplitude higher than 1 mmag is found in that region.
\subsection {Frequency and amplitude values}
There are two stars showing a frequency content clustering around a mean value,
i.e. HD 108100 (1.32 and 1.40 \cd) and $\gamma$ Dor itself (1.32, 1.36 and 
1.48 \cd);
on the other hand, HD 224945 and 9 Aur have at least one term largely separated
by the others. HR 2740 is very similar to the former stars, since there is
no indication of the presence of a term outside the interval 0.90--1.10 \cd.
However, the amplitudes are smaller than in the cases of 9 Aur and HD 108100;
moreover in HR 2740 there is a dominant term (1.04 \cd), while in those
cases the amplitudes are very similar. Hence, it is not possible to establish
common properties in the light curves of $\gamma$ Dor stars considering only
frequency and amplitude values.

Zerbi et al. (1997a; 1997b) emphasized the lack of amplitude in the $y$
light curve of 9 Aur and HD 164615 when
comparing, on the basis of theoretical models, the temperature ranges as
determined from the  different curves
$y, m_1, c_1, \beta$. This result finds a partial confirmation in the high
$A_v/A_y$ ratio announced by Breger et al. (1997) on the second frequency
of HD 101800 (the same ratio for the first frequency agrees with the
theoretical model). In the case of $\gamma$ Dor itself, the mean $A_b/A_y$ ratio
is around 1.20, for 9 Aur around 1.25, for HD 164615 around 1.30.
In the case of HR 2740, the $A_b/A_y$ ratio is around 1.26
for the $f_1, f_3, f_4$. Theoretical models of 
non-radial $p$ modes predict an $A_b/A_y$ ratio in the interval 1.20--1.30
(see, for example, Garrido et al. 1990). The theoretical model
of a $g$--mode pulsator proposed by Breger et al. (1997) does not change
appreciably the mean value of the $A_v/A_y$ ratio. Hence to establish how much 
the observed quantities deviate from theoretical predictions needs a more 
detailed investigation.

 As noted above, the $f_2$ term yields an $A_b/A_y$ ratio
close to 1.0, which can be explained by uncertainties on the zero-point
shifts, 
 rather than by a geometrical effect or by a surplus of $y$ amplitude.
Unfortunately, as often happens in the case of small amplitude variables, the
formal error bars reported in Tab. 3 precluded meaningful
evaluations of the phase shifts, which could supply some hints about mode
identifications.  
\subsection{Light curve stability}
The multiperiodicity of $\gamma$ Dor stars is a  well established fact on the
basis of the long term survey of $\gamma$ Dor itself and the results of the
recent multisite campaigns on 9 Aur and HD 224945. 
In those stars, after the subtraction of all known frequencies, the power spectra of 
the residuals often show a residual unexplained signal and a non-uniform
distribution of the noise around the mean light curve. In such a scenario,
the small amplitude variable HR 2740 is an exception, since the residual power
spectrum is flat (Fig. 5, bottom panel), the fit of the light curve quite
good (Fig. 4) and the scatter around the mean light curves is uniform (Fig. 6),
not supporting the possibility of a possible change in the maximum brightness
from one cycle to the next.
\subsection{The rotational modulation}
In many previous papers it was proposed to explain the complicated light
curve observed for the $\gamma$ Dor variables by the sum of one or two
close periodicities (eventually shaped as a double or triple wave) related to
the rotational period of the star.
The successive multisite campaignes provided much clearer power spectra and
it was realized that the observed light curves are generated by the sum of
a number of independent frequencies. The results obtained on HR 2740 and
their comparison with those obtained on  OU Pup and PR Pup (two Ap stars, i.e.
two rotating variables) described here yield other pieces of evidence against
a rotational variability. The light amplitude of HR 2740 (and of the other $\gamma$ Dor
stars) increases from $y$ to $v$, then decreases in $u$, as in the case of
pulsating stars; in the case of Ap
stars the amplitude is by far greater in $u$ than in other colours. Moreover,
the light curves of the Ap stars are monoperiodic and very stable in amplitude,
while the $\gamma$ Dor stars are multiperiodic and often have the
appearance of having  unstable amplitudes (Zerbi et al. 1997a for the 0.7679 \cd
 term in 9 Aur; Poretti et al. 1996
for all the terms in HD 224945). Of course, Ap stars are hotter  and maybe
the different physical conditions can introduce a completely different
behaviour, but it remains hard to explain why no term of the
$\gamma$ Dor light curves shows a flat part, as currently observed in a
rotating variable (see Fig. 2).
\subsection{The pulsational model}
Two attempts were made to yield a physical explanation of the observed 
frequencies, intepreted as different pulsation modes. Aerts \& Krisciunas 
(1996) discussed in detail the case of 9 Aur, demonstrating that the rotational
splitting can reasonably explain the observed two frequencies (the third one
was not known at that time); Breger et al. (1997) proposed a model of HD 108100
based on high--order, independent $g$--modes, without considering the effects
of the rotation. We performed some calculations using the procedure described
by Aerts \& Krisciunas (1996), demonstrating that in the case of HR 2740 it is
hard to explain the observed frequency separation ($<$0.20 \cd)  by the
splitting of different $m$ modes. Table 5 reports the
expected separation $\mid f-f_o\mid$ for $\ell$=1 and $\ell$=2, assuming a
first--order approximation, a $v\sin i$=40 km~s$^{-1}$ (Slettebak et al. 1975)
and $R=1.92R_\odot$. As can be noted, the frequency
spacing increases with $\ell$ and decreases with $i$, admitting a separation
of about 0.20 \cd only in the limiting case $\ell$=1 and $i>70^o$. However,
since most the observed terms have a smaller separation, the rotational
splitting does not appear suitable to match the observed frequency spectra of
HR 2740; this is also true considering the second order term since the 
$\Omega^2/f_o$ correction is marginal for $\Omega<0.5$ \cd and $f_o\sim$ 1 \cd).
Of course, this result should be considered with caution,
mindful of  the  warnings by Aerts \& Krisciunas (1996) about the
application of theoretical models of rotating $g$--mode pulsators.
\begin{table}
\centering
\caption{The rotational splitting $\mid f-f_o\mid$ (in \cd) calculated for
 $\ell$=1 and $\ell$=2 assuming $\mid m\mid$=1 and $v\sin i$=40 km~s$^{-1}$ 
 is shown in function of the $i$
angle (in degrees); $\Omega$ indicates the corresponding rotational frequency
(in \cd)}
\begin{tabular}{lrl ll}
\hline
\multicolumn{3}{c}{ }&\multicolumn{2}{c}{$\mid f-f_o\mid$}\\
\cline{4-5}
\multicolumn{1}{c}{$i$} & \multicolumn{1}{c}{$v_{eq}$} &
\multicolumn{1}{c}{$\Omega$} & \multicolumn{1}{c}{$\ell$=1}&
\multicolumn{1}{c}{$\ell$=2} \\
\noalign{\smallskip}
\hline
\noalign{\smallskip}
  10  &230&   2.37&   1.19& 1.97 \\
  20  &117&   1.20&   0.60& 1.00 \\
  30  & 80&   0.82&   0.41& 0.68 \\
  40  & 62&   0.64&   0.32& 0.53 \\
  50  & 52&   0.54&   0.27& 0.45 \\
  60  & 46&   0.48&   0.24& 0.39 \\
  70  & 42&   0.44&   0.22& 0.36 \\
  80  & 40&   0.42&   0.21& 0.35 \\
  90  & 40&   0.41&   0.21& 0.34 \\
\noalign{\smallskip}
\hline
\end{tabular}
\end{table}
The pulsation frequency spectrum of HR~2740 is reminiscent of the
very rich frequency spectra of $\delta$ Sct stars, where a large number of
independent frequencies were discovered by performing accurate multisite
campaigns (see for example the case of FG Vir, Breger 1995).
These campaigns gradually allowed asteroseismological models 
of the $\delta$ Sct stars to be proposed.
Perhaps this too can be the future of 
the rapidly developing field of $\gamma$ Dor stars.
\section*{Acknowledgements} 
D.~de~Alwis gratefully acknowledges the financial support of Dr.~Arthur 
C.~Clarke and the South African Astronomical Observatory which enabled her
attendance at the 1997 SAAO Summer School. She further thanks the Director
of the Arthur C.~Clarke Centre for Modern Technologies for leave of absence
during January 1997. P.~Martinez and C.~Koen acknowledge the technical 
support rendered by the SAAO workshops during these observations. E.~Poretti
thanks the ESO technical staff for the collaboration during the observations;
this project is one of the last carried out with the ESO 50-cm telescope
before its closing (1997 April 1).

\end{document}